\documentclass[prl,aps,twocolumn,superscriptaddress,tightenlines]{revtex4-2}
\usepackage{graphicx}
\usepackage{bm}
\usepackage{braket}
\usepackage{pifont}
\usepackage{amsmath}
\usepackage{amsfonts}
\usepackage{amssymb}    
\usepackage{CJKutf8}
\usepackage[english]{babel}
\usepackage{color,xcolor}
\usepackage[caption=false]{subfig}
\usepackage[normalem]{ulem}
\usepackage{natbib}
\usepackage{dsfont}
\usepackage{hyperref}
\hypersetup{
	colorlinks=true,
	linkcolor=blue,
	filecolor=black,  
	citecolor=blue, 
	urlcolor=blue,
	pdftitle={},
}

\newcommand{\phantomsubfloat}[1]{
	{% apply caption setup only temporarily
		\captionsetup[subfigure]{labelformat=empty}
		\subfloat[][]{#1}
	}%
}

\graphicspath{{./../Figs/}}
\begin{document}
\begin{CJK*}{UTF8}{bsmi}
\title{Efficient charging of multiple open quantum batteries through dissipation and pumping}
\author{Josephine Dias} 
\email{j.dias@uq.edu.au}
\thanks{These authors contributed equally}
\affiliation{Okinawa Institute of Science and Technology Graduate University, Onna-son$,$ Okinawa 904-0495$,$ Japan}
\affiliation{Centre for Quantum Computation and Communication Technology, School of Mathematics and Physics, University of Queensland, St Lucia, Queensland 4072, Australia}
\author{Hui Wang (王惠)} 
\email{huiwangph@gmail.com}
\thanks{These authors contributed equally}
\affiliation{Theoretical Quantum Physics Laboratory, Cluster for Pioneering Research, RIKEN, Wakoshi, Saitama 351-0198, Japan}
\author{Kae Nemoto}
\affiliation{Okinawa Institute of Science and Technology Graduate University, Onna-son$,$ Okinawa 904-0495$,$ Japan}
\affiliation{National Institute of Informatics, 2-1-2 Hitotsubashi, Chiyoda, Tokyo 101-0003, Japan}
\author{Franco Nori (野理)}
\affiliation{Theoretical Quantum Physics Laboratory, Cluster for Pioneering Research, RIKEN, Wakoshi, Saitama 351-0198, Japan}
\affiliation{Quantum Computing Center, RIKEN, Wakoshi, Saitama 351-0198, Japan}
\affiliation{Department of Physics, The University of Michigan, Ann Arbor, Michigan 48109-1040, USA}
\author{William J. Munro}
\affiliation{Okinawa Institute of Science and Technology Graduate University, Onna-son$,$ Okinawa 904-0495$,$ Japan}
\affiliation{National Institute of Informatics, 2-1-2 Hitotsubashi, Chiyoda, Tokyo 101-0003, Japan}
\date{\today}
\begin{abstract}
We explore a protocol that efficiently charges multiple open quantum batteries in parallel using a single charger. This protocol shows super-extensive charging through collective coupling of the charger and the battery to the same thermal reservoir. When applied to multiple quantum batteries, each coupled to different thermal reservoirs, the energy cannot be efficiently transferred from the charger to the battery via collective dissipation alone. We show that the counter-intuitive act of incorporating both dissipation and incoherent collective pumping on the charger enables efficient parallel charging of many quantum batteries.
\end{abstract}
\maketitle
\end{CJK*}

Quantum batteries (QB) have attracted significant recent attention for the potential of transferring and storing energy via quantum mechanically enhanced processes \cite{alicki2013entanglement, binder2015quantacell, Skrzypczyk2014,PhysRevLett.111.240401, PhysRevA.97.022106, Crescente_2020, PhysRevLett.122.210601, PhysRevA.101.032115, RevModPhys.96.031001}. Such devices could prove useful in the control, manipulation and powering of quantum technologies. Initially, QBs were considered in a closed system with unitary dynamics, where it was demonstrated that using entangling unitary operations allows for the extraction of more work \cite{alicki2013entanglement}. QBs have since been considered in several open system frameworks \cite{farina2019charger, liu2019loss}. In particular, the Dicke model \cite{dicke1954coherence}, which considers an ensemble of spins collectively coupled to a photon mode, can be employed to speed up the charging time of a QB \cite{ferraro2018high, PhysRevB.102.245407, dou2022extended, yang2024resonator, carrasco2022collective, zhang2023enhanced}. The crucial aspect is that quantum characteristics---such as entanglement, quantum coherence, and superradiant (collective) behaviours---can enable more powerful and efficient charging processes compared to classical methods \cite{binder2015quantacell, campaioli2017enhancing}. 

Recently, an open quantum battery protocol was introduced where a charger, initially comprising an ensemble of spins in the excited state, charges a battery consisting of a smaller spin ensemble \cite{quach2020using}. The charger and battery are not directly coupled to each other, but are collectively coupled to a single thermal reservoir. The charging process is initiated by switching on the coupling between the spin ensembles and the reservoir. Collective dissipation in the entire system leads to spin inversion \cite{hama2018negative, hama2018relaxation}, thereby charging the battery. Counter-intuitively, the energy stored in the spins of the battery does not decay to the ground state even in the presence of a zero-temperature thermal reservoir: the charged state of the battery remains stable while connected to the charger. This phenomenon arises because the steady state of the charger and battery includes a component that does not interact with the reservoir and thus avoids decaying to the ground state. Such states are known as dark states, which emerge when multiple spins are collectively coupled to a bosonic mode \cite{doi:10.1139/p67-142, PhysRevA.5.1094}. This is due to the collective coupling of all the spins in the battery and charger to the thermal reservoir and should not be viewed as energy transfer due to the emission of a photon by the charger followed by the absorption of that photon by the battery. 

Provided the number of spins in the charger is significantly larger than the number of the spins in the battery, all the spins in the battery will transition to the excited state \cite{quach2020using}. As part of our research, we take the next critical step to \textit{simultaneously charge multiple batteries in parallel, each located in separate reservoirs using a single, centralized charger.} Our method for efficiently charging multiple quantum batteries incorporates external resources to continuously pump the charger, thereby providing a steady supply of energy to the charger-QB system. This protocol relies on multiple incoherent processes including collective dissipation and pumping.

We start by exploring the model of a general charger-battery system, consisting of two ensembles of identical spin$-1/2$ particles with frequency $\omega_0/2\pi$. Both ensembles are collectively coupled to the same bosonic reservoir. One ensemble is designated as the charger $C$, and the other as the battery $B$. The Hamiltonian for this system is:
\begin{equation}
\begin{split}
H=&\hbar \omega_0  (J^z_C+J_B^z) +\sum_k \hbar \omega_k a^\dagger_k a_k
\\
& + \sum_k \left[ \lambda_k \left( J^+_C+J^+_B  \right)a_k +\lambda^*_k a^\dagger_k \left( J_C^-+J_B^- \right)  \right] .
\end{split}
\label{eq:ham}
\end{equation}
Here, the first term represents the spins in the QB and charger, the second term is the Hamiltonian of the bosonic reservoir, the third term represents the collective interaction of all spins in the QB and charger with the bosonic modes, while $J_B,J_C$ are the collective spin operators for the battery and charger respectively. These are defined as $J_{C,B}^\alpha=1/2 \Sigma_{i=1}^{N_C,N_B} \sigma_i^\alpha $, with $\alpha=x,y,z$, where $\sigma_i^\alpha$ is the Pauli operator of the $i^{\mathrm{th}}$ spin. The collective spin raising and lowering operators are $J_{C,B}^{\pm}= J_{C,B}^x\pm i J_{C,B}^y$. The operators $a_k^\dagger $ ($a_k$) represent the creation (annihilation) operators of the $k^{\mathrm{th}}$ bosonic mode of the reservoirs. The parameters $\lambda_k$ ($\lambda_k^*$) represent emission (absorption) amplitudes that fix the spectral density of the reservoirs $\Gamma(\omega) = 2\pi \sum_k |\lambda_k|^2 \delta(\omega - \omega_k)$. Turning on or off the collective coupling will start or stop the charging process.

Under the Born-Markov approximation, in the weak-coupling limit, the master equation describing the dynamics of the combined system can be written in the rotating frame as:
\begin{equation}
	\begin{split}
		\dot\rho =& \gamma_{\Downarrow} \mathcal{L}  \left[ J^-_C+J^-_B\right]\rho \ ,
	\end{split}
	\label{eq:oneQB}
\end{equation}
for a zero-temperature reservoir. Here, $\gamma_{\Downarrow}$ is the dissipative collective coupling constant between spin-ensembles and reservoir with $\gamma_{\Downarrow}=\Gamma(\omega_0)=\Gamma \omega_0 $ in the wide band limit. The superoperator $\mathcal{L} $ is defined by $\mathcal{L} \left[O\right] \rho = 2 O \rho O^\dagger -O^\dagger O \rho - \rho O^\dagger O$. Next the energy density of the spin ensembles is defined as:
% \begin{equation}
% \frac{\mathcal{E}_{C/B} (t)}{\hbar\omega_0} =  \frac{\braket{J_{C/B}^z (t)}}{N_{C/B}}+ \frac{1}{2} \ ,
% \end{equation}
\begin{equation}
\frac{\mathcal{E}_{\mu} (t)}{\hbar\omega_0} =  \frac{\braket{J_{\mu}^z (t)}}{N_{\mu}}+ \frac{1}{2} \ ,
\end{equation}
 where the subscript $\mu$ denotes the charger $C$ or battery $B$. Initially, the energy densities of the system are $\mathcal{E}_C \left(0\right)/(\hbar\omega_0)=1$ and $\mathcal{E}_B \left(0\right)/(\hbar\omega_0)=0$. Efficient transfer means the energy density of the battery will go to $\mathcal{E}_B \left(t_\mathrm{f}\right)/(\hbar\omega_0)\sim 1$ at some time $t_\mathrm{f}$. 
 
Nominally, there are two scales for which we can look at this open quantum battery protocol \cite{quach2020using}. Small scale systems for which the number of spins in the batteries and the charger are kept small, these can be modeled via solving the master equation directly. In systems where the battery and the charger have large spin numbers, solving the master equation exactly is challenging. Here, we can employ a mean-field approximation to solve for the dynamics at large scales. 

In general, for this process to transfer energy from the charger to the QB [$\mathcal{E}_B(t_{\mathrm{f}})>\mathcal{E}_C(t_{\mathrm{f}})$], it requires the condition $N_C > N_B$; that is, the charger must be larger than the battery. Efficient energy transfer [$\mathcal{E}_B/(\hbar\omega_0)\to 1$] occurs when $N_C \gg N_B$. Notably, this charging process is most efficient for zero-temperature reservoirs. However, as the reservoir temperature increases, energy transfer can still be achieved, albeit the process  is less efficient [$\mathcal{E}_B(t_{\mathrm{f}})/(\hbar\omega_0)< 1$]. Importantly, while the spins in the battery and the charger are not directly coupled, the collective coupling between all spins and the thermal reservoir allows for effective long range interactions between the battery and the charger. Entanglement is generated between the battery and the charger through this process \cite{hama2018negative,dias2023entanglement}.

\begin{figure}[t!]
	\centering
	\includegraphics[width=0.9\linewidth]{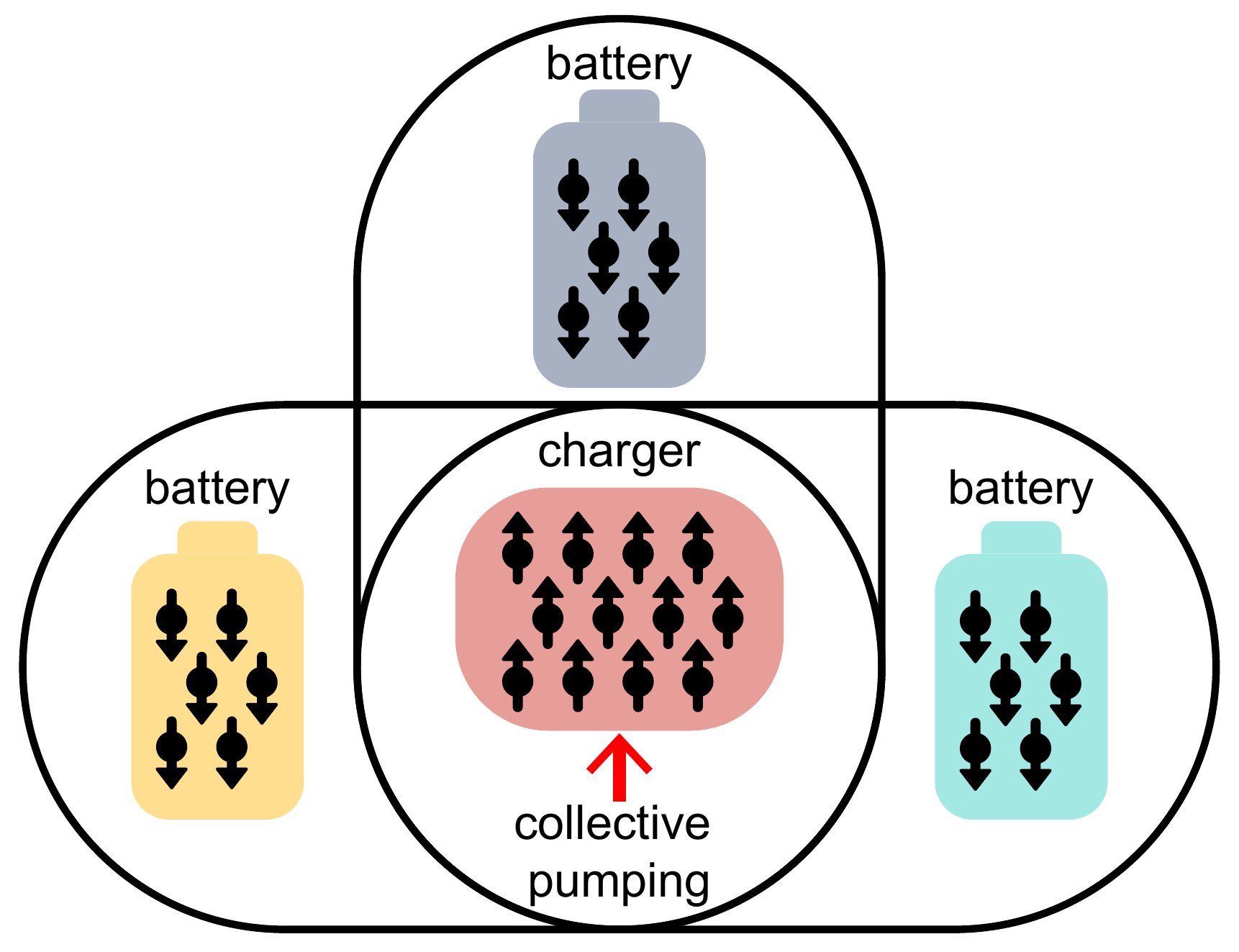}
	\caption{Open quantum battery protocol for efficient charging of many QBs coupled to separate thermal reservoirs. A central spin ensemble labelled `charger' is coupled to three separate thermal reservoirs. Each individual reservoir contains a QB (a spin ensemble smaller than the charger). Pictured here is the initial state of the total system, with all spins in the charger prepared in the excited state and all spins in all QBs in the ground state. Through the combination of the collective spin relaxation and the collective pumping on the charger, the energy density can be transferred from the charger to the battery. }
	\label{fig:battery}
\end{figure}

In the following, we explore how to utilize this process to charge multiple QBs in parallel. While it is known that many QBs can be efficiently charged if they are coupled to the same reservoir \cite{ferraro2018high}, this induces a severe spatial constraint on the location of the QBs, as they must all couple to the same field. Ideally, we would want to utilize a central charger that can be coherently coupled to many different thermal reservoirs, with each reservoir holding its own QB. This allows the QBs to be potentially further apart and can in principle direct energy towards other devices. Another advantage of coupling to independent reservoirs could be the ability to accommodate asymmetrical coupling configurations~\cite{mojaveri2024extracting}.

\begin{figure*}
\centering \phantomsubfloat{\label{fig:qb2}}\phantomsubfloat{\label{fig:qb3}} \phantomsubfloat{\label{fig:qb3drive}} \phantomsubfloat{\label{fig:qb4}} \phantomsubfloat{\label{fig:qb4driveG1}}  \phantomsubfloat{\label{fig:qb4drive}}
\includegraphics[width=0.99\linewidth]{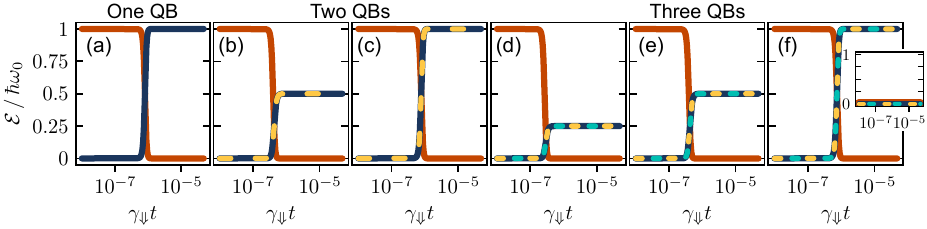}
	\caption{Energy density charging QBs with collective dissipation and pumping. The red curve corresponds to the `charger' with a spin population of $N_C=10^7$. The three batteries correspond to the solid dark blue, dashed yellow and dashed green curves, each with the same spin population of $N_{B_1}=N_{B_2}=N_{B_3}=10^2$. The battery curves of each plot all overlap. In \protect\subref{fig:qb2}, the charger is coupled to a single reservoir with one QB using collective dissipation and no collective pumping ($\gamma_\Uparrow=0$). In~\protect\subref{fig:qb3},  the charger is coupled to two reservoirs each holding a QB of $N_B=10^2$ spins and no collective pumping is applied to the charger ($\gamma_\Uparrow=0$). In~\protect\subref{fig:qb3drive}, the system is the same as \protect\subref{fig:qb3}, but an incoherent collective pump of strength $\gamma_\Uparrow=\gamma_\Downarrow$ is now applied to the charger. In \protect\subref{fig:qb4}, the charger is coupled to three reservoirs each holding a QB and no collective pumping is applied to the charger ($\gamma_\Uparrow=0$). In~\protect\subref{fig:qb4driveG1}, the system is the same as~\protect\subref{fig:qb4}, but an incoherent collective pump of strength $\gamma_\Uparrow=  \gamma_\Downarrow$ is now applied to the charger. In~\protect\subref{fig:qb4drive}, the system is the same as~\protect\subref{fig:qb4driveG1}, with an incoherent collective pump of strength $\gamma_\Uparrow= 2 \gamma_\Downarrow$. The inset shows dynamics with collective pumping ($\gamma_{\Uparrow} =2 \gamma _{\Downarrow}$) but with $\mathcal{E}_C(0)=0$, indicating that it is both the collective spin relaxation and collective pumping combined that facilitates the charging of these QBs.}
 \label{fig:results}
\end{figure*}

Extending the model from a single QB in one reservoir \eqref{eq:oneQB} to multiple QBs each in its own separate reservoir, yields the following master equation:

\begin{equation}
	\begin{split}
		\dot\rho =&\gamma_{\Downarrow} \sum_m \mathcal{L}  \left[ J^-_C+J^-_{B_m}\right]\rho  + \gamma_{\Uparrow} \mathcal{L} \left[J^+_{C}\right] \rho \ ,
	\end{split}
	\label{eq:nodrive}
\end{equation}
where the presence of $m$ zero-temperature reservoirs yields incoherent dissipative processes between the central charger and the $m$ QBs. This system is depicted in Fig.~\ref{fig:battery} for $m=3$. The last term in \eqref{eq:nodrive} corresponds to incoherent collective pumping on the charger with $\gamma_\Uparrow$ representing the rate of collective pumping. We use a mean-field approximation on \eqref{eq:nodrive} to model the system. The energy density dynamics of the charger and the QBs is given in  Fig.~\ref{fig:results}.  First, in Fig.~\ref{fig:qb2}, we present the large-scale dynamics of the charger and a single QB from the original proposal in Ref.~\cite{quach2020using}. This scheme does not involve collective pumping ($\gamma_\Uparrow=0$). This dynamics shows the charger being initialized with $\mathcal{E}_C(0)/(\hbar \omega_0 )=1$, corresponding to all spins in the charger being in the excited state. Here, we can see efficient charging showing $\mathcal{E}_B/(\hbar \omega_0)\to 1$, and this happens on a superradiant time-scale, demonstrating the collective speed up due to the presence of many spins. In Figs.~\ref{fig:results}~(b, c), the charger is coupled to two reservoirs each holding a QB of $N_B=10^2$ spins. In Figs.~\ref{fig:results}~(d - f), the charger is coupled to three reservoirs each holding a QB. Figs.~\ref{fig:results}~(b, d) illustrate what happens when this system is extended to charge more batteries in more reservoirs without collective pumping. The steady-states of the two [Fig.~\ref{fig:qb3}] and three [Fig.~\ref{fig:qb4}] QB systems reveal a crucial limitation when this scheme is extended to multiple thermal reservoir couplings. While only one specific spin population configuration is shown here, through numerical exploration, for two reservoirs, we find that with no collective pumping, the limit of $\mathcal{E}_B/(\hbar \omega_0)\to 1/2$ cannot be surpassed no matter how much larger $N_C$ is than $N_B$. For three reservoirs, this limit is further reduced to $\mathcal{E}_B/(\hbar \omega_0) \to 1/4$. It is important to stress that this effect is due to the presence of additional decay channels through the coupling to multiple thermal reservoirs. If the three QBs were all contained in the same reservoir with the charger, one could witness efficient charging. However, the presence of multiple thermal reservoirs severely diminishes the charging capability of this protocol using dissipation alone.

In the case of a single thermal reservoir, collective dissipation is sufficient to achieve high power collective charging of a QB. However, when this model is extended to multiple reservoirs with multiple QBs, the presence of these decay channels mean that many QBs cannot be efficiently charged in parallel via the same mechanism. Therefore, one might conclude that charging multiple QBs through multiple collective dissipation channels is not a viable strategy. To address this, we will now introduce the main result of this work. Here, we present an adapted strategy to efficiently charge multiple QBs in parallel.  Our protocol is pictured schematically in Fig.~\ref{fig:battery} and depicts a central spin ensemble labeled as the `charger' coupled to three separate thermal reservoirs. Coupled to each individual reservoir is its own QB. The initial state of the system before charging the QBs has the charger initialized in the fully excited state and all QBs in their ground state. To efficiently charge all three QBs in parallel, continuous collective pumping must be applied to the central charger as the system simultaneously experiences collective dissipation. Results are given in Figs.~\ref{fig:results}~(c, f). Here, we see that through the addition of a incoherent collective pump on the charger only, we can achieve efficient energy transfer from charger to battery. By comparison with Fig.~\ref{fig:qb2}, our scheme can produce the same highly efficient energy transfer, from charger to multiple batteries in separate reservoirs. \textit{We empahsize that this effect cannot be seen if the charger is coherently pumped. It is the presence of the incoherent collective pump that facilitates efficient energy transfer. }

From Fig.~\ref{fig:results}, we observe that the presence of the incoherent pump slightly slows the superradiant decay when the number of QBs is fixed. While the pump adds more energy, the entire system takes longer to relax, thereby slightly extending the charging time. Nonetheless, the charging process still occurs on the timescales of collective effects. Lastly, while entanglement can be generated between spins in different spin ensembles (QBs and the charger) in the absence of pumping  \cite{dias2023entanglement}, we find that with collective pumping, there is little to no entanglement between the QBs and the charger. Despite the absence of entanglement, we find efficient energy transfer between a charger and multiple QBs. We note it has recently been shown \cite{zhang2023enhanced} that entanglement is not associated with the collective charging speed up in the Dicke quantum battery, rather the enhancement is due to coherent cooperative interactions.  

The collective pumping strength $\gamma_{\Uparrow}$ required for efficient charging depends on the number of thermal reservoirs coupled to the charger. In Fig.~\ref{fig:results}~(d, e, f), we focus on the three reservoir system (depicted in Fig.~\ref{fig:battery}) and show the dynamics for pumping strength $\gamma_{\Uparrow}=\gamma_{\Downarrow}$ in Fig.~\ref{fig:qb4driveG1}. We observe that $\gamma_{\Uparrow}=\gamma_{\Downarrow}$ can only half excite the QBs, and $\gamma_{\Uparrow}=2\gamma_{\Downarrow}$ [Fig.~\ref{fig:qb4drive}] is required to fully charge the three QBs in the system. In principle, our strategy could be adapted to charge more QBs from one single centralized charger. However, we expect that as more reservoirs are utilized, higher collective pumping strengths are likely needed for efficient charging in this system. The ergotropy of this quantum battery protocol has already been considered for the single QB system \cite{quach2020using}.

Our scheme can be implemented in many different hybrid quantum systems \cite{xiang2013hybrid}, including spin ensembles in the microwave regime coupled to waveguide or microwave resonators, which act as reservoirs. A possible implementation could utilize diamond samples with nitrogen vacancy defect centers as the spin ensembles \cite{angerer2016collective, angerer2018superradiant}. In this system, coherent coupling of two distinct spin ensembles has been realized \cite{astner2017coherent}. Additionally, another candidate for the spin ensembles could be erbium dopants, which operate in the optical region \cite{PhysRevX.10.041025}.

Lastly, we remark on a few key aspects of this work. While the dynamics presented does not include individual effects such as individual decay, we emphasize that these effects occur on timescales that are much slower than collective effects for large ensemble sizes. Further, one may wonder if superradiant decay of the charger is necessary for efficient energy transfer in our scheme with collective pumping. To address this, we refer to the inset in Fig.~\ref{fig:qb4drive}. This inset shows the energy density dynamics following \eqref{eq:nodrive} with $\mathcal{E}_C(0)=0$. On comparable timescales, no transfer of energy is observed and therefore we stress that it is the combination of both the spin relaxation and collective pumping that enables this charging process. 

 In conclusion, we have introduced a method that circumvents limitations induced on charging multiple quantum batteries in separate reservoirs through collective dissipation alone. By performing incoherent, collective pumping on the charger, we can fully charge many QBs while they are indirectly coupled to the charger through independent reservoirs. This protocol permits the QBs to be somewhat spatially separated as they do not all need to be coupled to the same reservoir. Our method can be scaled to charge more batteries than shown here, although a stronger collective pump is likely needed in such a case. This protocol can also be applied to transfer energy and quantum correlations between multiple spin ensembles arranged in series \cite{dias2021reservoir, munro2021collective, dias2023entanglement}. Our results have important implications of how the flow of energy can be manipulated in quantum devices. This paves the way for advances in quantum thermodynamics and outlines potential directions for quantum technologies. 

 We thank Neil Lambert, Dario Poletti, and Christopher W. Wächtler for fruitful discussions. This work was supported by the MEXT Quantum Leap Flagship Program (MEXT QLEAP) Grant No. JPMXS0118069605. This research was partially supported by the Australian Research Council Centre of Excellence for Quantum Computation and Communication Technology (Project No. CE170100012). F.N. is supported in part by: Nippon Telegraph and Telephone Corporation (NTT) Research, the Japan Science and Technology Agency (JST) [via the CREST Quantum Frontiers program Grant No. JPMJCR24I2, the Quantum Leap Flagship Program (Q-LEAP), and the Moonshot R\&D Grant Number JPMJMS2061], and the Office of Naval Research (ONR) Global (via Grant No. N62909-23-1-2074).

\end{document}